\documentstyle[12pt]{article}
\newcommand{\bm}[1]{\mbox{\boldmath$#1$}}
\def\ds{\displaystyle}
\begin{document}

\centerline{\large{\bf  A Characterization of Discrete Time Soliton Equations}}

\vglue 1cm

\centerline{{Satoru SAITO$^*$ \footnote{E-mail: saito@phys.metro-u.ac.jp},\quad Noriko SAITOH$^{**}$\footnote{E-mail: saitoh@lam.osu.sci.ynu.ac.jp}, }}
\centerline{Jun-ichi YAMAMOTO$^*$\footnote{E-mail: yjunichi@kiso.phys.metro-u.ac.jp} and\quad Katsuhiko YOSHIDA$^{***}$\footnote{E-mail: yoshida@kiso.phys.metro-u.ac.jp}}
\vglue 1cm
\centerline{${}^{*}${\it Department of Physics, Tokyo Metropolitan University}}
\centerline{{\it Minamiohsawa 1-1, Hachiohji, Tokyo 192-0397 Japan}}
\centerline{${}^{**}${\it Department of Applied Mathematics, Yokohama National University}}
\centerline{\it Hodogaya-ku, Yokohama, Japan 240-8501}
\centerline{{\it ${}^{***}$School of Science, Kitasato University, 1-15-1 Kitasato}}
\centerline{{\it Sagamihara, Kanagawa, 228-8555 Japan}}
\vglue 1cm
\centerline{\bf Abstract}
\begin{center}\begin{minipage}{14cm}{\small{We propose a method to characterize discrete time evolution equations, which generalize discrete time soliton equations, including the $q$-difference Painlev\'e IV equations discussed recently by Kajiwara, Noumi and Yamada\cite{KNY}.}}
\end{minipage}\end{center}

\sloppy
\section{Introduction}

A discretization of independent variables of integrable nonlinear differential equations breakes their integrability in general. Therefore it is remarkable that there exist certain discrete analogue of integrable differential equations which preserve integrability\cite{Hirota}$^-$\cite{Su}. Integrability of ODEs can be tested by studying whether exist singularities which depend on initial values, a method called Painlev\'e test. Therefore it is natural to look for a discrete analogue of the Painlev\'e property which enables one to predict behaviour, deterministic or chaotic, of a sequence of map starting from some initial value. There has been, however, not known a discrete analogue of Painlev\'e test despite of some useful proposals\cite{HV}. 

In our previous paper we have studied a discrete analogue of the Lotka-Volterra equation which is known completely integrable\cite{NSSY}. Considering it as a sequence of a map it was observed that at every step of the map there are two possibilities to be chosen. Hence the orbit is not deterministic. Nevertheless the map is compatible with integrability of the equation. This happens due to the fact that one of two types of the map, which we call B-type, does not generate a new orbit but simply exchanges two different orbits created by the other type (A-type) of the map.

Starting from a symmetric discrete Lotka-Volterra equation (dLV) under the periodic boundary conditions the map is determined solving a quadratic equation. The expression under the square root which appears in the solutions turns out to be a perfect square in this problem, hence no singularity arises in the map. Out of the two maps the A-type map becomes one which reduces to the ordinary Lotka-Volterra equation in the continuous time limit, whereas the B-type map has no correspondence in the continuous time. If the equation is deformed arbitrary, the square root singularity remaines and one can not obtain regular map.

We now ask if the disappearance of the square root singularity is special in the dLV. The purpose of this paper is to show that the phenomena is quite general among various discrete integrable systems including discrete KdV, discrete time Toda and discrete KP equations as well as their $q$-difference versions.

In order to clarify our problem we review briefly, in the following section, the argument of the dLV in our previous paper\cite{NSSY}. We generalize the dLV in \S 3 and show that the same argument holds in this case too. This generalization includes, as a special case, the $q$-difference fourth Painlev\'e equation discussed recently in ref\cite{KNY}. We can further generalize the scheme including arbitrary functions which we discuss in \S 4. $q$-difference discrete KdV, Toda, KP equations will be obtained as special reductions. In the last section conservation laws are discussed.

\section{Discrete Lotka-Volterra equation}

The discrete time Lotka-Volterra system is defined by the set of equations
\begin{equation}
x_n^{j+ \delta}(1- \delta x_{n-1}^{j+\delta})=x_{n}^j(1-\delta x_{n+1}^j),\qquad j\in \delta \bm{Z},\quad n=1,2,\cdots, N
\end{equation}
together with the boundary conditions
$$
x_{N+1}^j=x_1^j,\qquad j\in \delta \bm{Z}.
$$
Here $\delta$ denotes the minimum interval of the time step. This system is known completely integrable\cite{HT}. The integrability is justified by noting that it can be obtained from the discrete time Toda lattice via a Miura transformation, or existence of sufficient number of conserved quantities. 

We have investigated behaviour of solutions to this equation and found two types of map, one of which has a correspondence in the continuous time limit and the other does not\cite{NSSY}. To be more precise let us consider the case of $N=3$, and write the equations as 
\begin{equation}
\cases{
\bar x(1-  \bar z)=x(1-  y)\cr
\bar y(1-  \bar x)=y(1-  z)\cr
\bar z(1-\bar y)=z(1-x),\cr}
\label{3dLV}
\end{equation}
where we use the notations 
\begin{equation}
(x,y,z)=(\delta x_{1}^{j},\delta x_{2}^{j},\delta x_{3}^{j}),\qquad (\bar x, \bar y, \bar z)=(\delta x_1^{j+\delta},\delta x_2^{j+\delta},\delta x_3^{j+\delta}).
\label{x,y,z}
\end{equation}

In order to clarify our problem let us consider the following equations at first
\begin{equation}
\bar x(1-\bar z)=X,\qquad \bar y(1-\bar x)=Y,\qquad \bar z(1-\bar y)=Z.
\label{X,Y,Z}
\end{equation}
Here $(X,Y,Z)$ are some functions of $(x,y,z)$. Solving (\ref{X,Y,Z}) for $\bar x$, we find
\begin{equation}
\bar x= {1+X-Y-Z\pm\sqrt{(1-X-Y-Z)^2-4XYZ}\over 2(1-Z)},
\label{bar x=}
\end{equation}
and similarly for $\bar y$ and $\bar z$.

This result shows that the mapping is not single valued. Hence there arises naturally a question, how this double valuedness is compatible with the integrable nature of the Lotka-Volterra equation (\ref{3dLV}). This problem is answered if we substitute the right hand sides of (\ref{3dLV}) into $(X,Y,Z)$ of (\ref{bar x=}). In fact we obtain
\begin{equation}
\left(\matrix{\bar x\cr \bar y\cr \bar z\cr}\right)
=\cases{
\left(\matrix{
x\ds{{1- y+yz\over 1- z+xz}}\cr
y\ds{{1-z+zx\over 1-x+yx}}\cr
z\ds{{1-x+xy\over 1-y+zy}}\cr}
\right)\qquad  {\rm : A-type}\cr
\qquad\cr
\qquad\cr 
\left(\matrix{1-y\cr 1-z\cr 1-x\cr}\right)\qquad
{\rm : B-type.}\cr}
\label{3dLV solution}
\end{equation}
Note that the B-type map does not generate orbits but simply exchanges the variables. If we operate B six times, for instance, all points return to the original place after the sequence of the map:
$$
(x,y,z)\rightarrow (1-y,1-z,1-x)\rightarrow (z,x,y)\rightarrow (1-x,1-y,1-z)
$$
$$
\qquad\qquad \rightarrow (y,z,x)\rightarrow (1-z,1-x,1-y)\rightarrow (x,y,z).
$$

There exist two constants of the equation of motion (\ref{3dLV}):
\begin{eqnarray}
C&=&x+y+z-xy-yz-zx
\label{C}\\
D&=&xyz(1-x)(1-y)(1-z).
\label{D}
\end{eqnarray}
They are conserved by both A- and B-type of maps. Moreover we also found that there are two constants
\begin{equation}
r=xyz,\qquad s=(1-x)(1-y)(1-z)
\label{r,s}
\end{equation}
which are conserved under the A-type map but are exchanged by the B-type map:
\begin{eqnarray*}
(r,s)&\Leftrightarrow &(r,s)\qquad {\rm for\ A-type},\\
(r,s)&\Leftrightarrow &(s,r)\qquad {\rm for\ B-type}.
\end{eqnarray*}

In order to see the continuous time limit of the map we must recover $\delta$ in the old variables (\ref{x,y,z}). We will find that in the limits $\delta\rightarrow 0,\ j\rightarrow t$, the equations (\ref{3dLV}) as well as the A-type map of (\ref{3dLV solution}) reduce to the symmetric Lotka-Volterra equations
\begin{equation}
{dx\over dt}=x(z-y),\qquad
{dy\over dt}=y(x-z),\qquad
{dz\over dt}=z(y-x).
\label{LV equation}
\end{equation}
On the other hand the B-type map does not have a proper limit as it was discussed in \cite{NSSY}. 


\section{$q$-Lotka-Volterra maps}

The integrability of the Lotka-Volterra map (\ref{3dLV}) owes to its special symmetry under the cyclic permutation of variables. If we replace it by, for instance,
\begin{equation}
\cases{
\bar x(1-\bar z)=ax(1-y)\cr
\bar y(1-\bar x)=az(1-z)\cr
\bar z(1-\bar y)=az(1-x),\cr}
\label{3dLV with a}
\end{equation}
the integrability is lost when $a\ne 1$. Therefore it is highly nontrivial question whether there exists generalization of the discrete Lotka-Volterra system (\ref{3dLV}) which preserves integrability. 

We consider in this section the following set of equations as a generalization of (\ref{3dLV}):
\begin{equation}
\cases{
\bar x(1-\bar z/c)=aX,\cr
\bar y(1-\bar x/a)=bY,\cr 
\bar z(1-\bar y/b)=cZ,\cr}
\label{3dGLV}
\end{equation}
where $(X,Y,Z)$ are some functions of $(x,y,z)$ and $a,b,c$ are arbitrary constants. By the same reason we discussed in the previous section, we expect maps of the form (\ref{bar x=}) when we solve (\ref{3dGLV}) for $(\bar x,\bar y,\bar z)$ in general. In fact we obtain
\begin{equation}
\bar x=a{1+X-Y-Z\pm\sqrt{(1-X-Y-Z)^2-4XYZ}\over 2(1-Z)}.
\label{square root}
\end{equation}

Motivated by our previous experience we try to find $(X,Y,Z)$ such that the expression under the square root is a perfect square. We found the following three cases satisfy this condition:

\begin{equation}
\cases{
\bar x(1-\bar z/c)=a^2x(1-by),\cr
\bar y(1-\bar x/a)=b^2y(1-cz),\cr
\bar z(1-\bar y/b)=c^2z(1-ax),\cr}
\label{X=x(1-y)}
\end{equation}
\begin{equation}
\cases{
\bar x(1-\bar z/c)=aby(1-cz),\cr
\bar y(1-\bar x/a)=bcz(1-ax),\cr
\bar z(1-\bar y/b)=cax(1-by),\cr}
\label{X=y(1-z)}
\end{equation}
\begin{equation}
\cases{
\bar x(1-\bar z/c)=acz(1-ax),\cr
\bar y(1-\bar x/a)=bax(1-by),\cr
\bar z(1-\bar y/b)=cby(1-cz).\cr}
\label{X=z(1-x)}
\end{equation}

Solving them for $(\bar x,\bar y,\bar z)$ we obtain
\begin{equation}
\left(\matrix{\bar x\cr \bar y\cr \bar z\cr}\right)
=
\cases{
\left(\matrix{a^2x\ds{{1-by+bcyz\over 1-cz+acxz}}\cr 
b^2y\ds{{1-cz+acxz\over 1-ax+abxy}}\cr c^2z\ds{{1-ax+abxy\over 1-by+bcyz}}\cr}\right)\qquad :{\rm A-type}
\cr
\ \ \ \ 
\cr
\ \ \ \ 
\cr
\left(\matrix{a(1-by)\cr b(1-cz)\cr c(1-ax)\cr}\right)\qquad :{\rm B-type}
}
\label{17}
\end{equation}
for the first set,
\begin{equation}
\left(\matrix{\bar x\cr \bar y\cr \bar z\cr}\right)
=
\cases{
\left(\matrix{aby\ds{{1-cz+caxz\over 1-ax+abxy}}\cr 
bcz\ds{{1-ax+abxy\over 1-by+bcyz}}\cr acx\ds{{1-by+bcyz\over 1-cz+acxz}}\cr}\right)\qquad :{\rm A-type}
\cr
\ \ \ \ 
\cr
\ \ \ \ 
\cr
\left(\matrix{a(1-cz)\cr b(1-ax)\cr c(1-by)\cr}\right)\qquad :{\rm B-type}
}
\label{maps of the second set gdLV}
\end{equation}
for the second set, and
\begin{equation}
\left(\matrix{\bar x\cr \bar y\cr \bar z\cr}\right)
=
\cases{
\left(\matrix{acz\ds{{1-ax+abxy\over 1-by+bcyz}}\cr 
abx\ds{{1-by+bcyz\over 1-cz+acxz}}\cr bcy\ds{{1-cz+acxz\over 1-ax+abxy}}\cr}\right)\qquad :{\rm A-type}
\cr 
\ \ \ \ 
\cr
\ \ \ \ 
\cr
\left(\matrix{a(1-ax)\cr b(1-by)\cr c(1-cz)\cr}\right)\qquad :{\rm B-type}
}
\end{equation}
for the last set.

Note that the first set (\ref{X=x(1-y)}) is the Lotka-Volterra map (\ref{3dLV solution}) when $a=b=c=1$. The A-type map (\ref{maps of the second set gdLV}) of the second set is the $q$-Painlev\'e IV studied by Kajiwara, Noumi and Yamada\cite{KNY}.

From these expressions it is apparent that the above three cases are symmetric under the cyclic permutations:
\begin{equation}
(ax,by,cz)\rightarrow (by,cz,ax) \rightarrow (cz,ax,by).
\label{cyclic permutation}
\end{equation}
We will call this symmetry the cyclic symmetry. Owing to this symmetry it is sufficient to consider only one of three equations.

In addition to the cyclic symmetry, the symmetry under the exchange of the variables 
$$
(\bar x/a,\bar y/b,\bar z/c)\leftrightarrow (ax,by,cz)
$$
in equations (\ref{X=x(1-y)}),(\ref{X=y(1-z)}),(\ref{X=z(1-x)}) enables us to solve them for $(x,y,z)$ in terms of $(\bar x,\bar y,\bar z)$. We call this symmetry the time reversal symmetry. For instance we solve, corresponding to (\ref{X=x(1-y)}),
\begin{equation}
\cases{
a^2x(1-cz)=\bar x(1-\bar y/b),\cr
b^2y(1-ax)=\bar y(1-\bar z/c),\cr
c^2z(1-by)=\bar z(1-\bar x/a),\cr}
\label{bar x(1-bar y)=x(1-z)}
\end{equation}
for $(x,y,z)$ and obtain
$$
\left(\matrix{x\cr y\cr z\cr}\right)
=
\cases{
\left(\matrix{\ds{{\bar x\over a^2}}\ds{{1-\bar z/c+\bar y\bar z/bc\over 1-\bar y/b+\bar x\bar y/ab}}\cr 
\ds{{\bar y\over b^2}}\ds{{1-\bar x/a+\bar x\bar z/ac\over 1-\bar z/c+\bar y\bar z/bc}}\cr \ds{{\bar z\over c^2}}\ds{{1-\bar y/b+\bar x\bar y/ab\over 1-\bar x/a+\bar x\bar z/ac}}\cr}\right)\qquad :{\rm A-type}
\cr
\ \ \ \ 
\cr
\ \ \ \ 
\cr
\left(\matrix{(1-\bar z/c)/a\cr (1-\bar x/a)/b\cr (1-\bar y/b)/c\cr}\right)\qquad :{\rm B-type}.
}
$$
We could obtain this result by solving (\ref{17}) for $(\bar x,\bar y,\bar z)$.

Generalization of our argument to $N>3$ is straightforward. For example the 4-point discrete Lotka-Volterra equation is generalized to
\begin{equation}
\cases{
\bar x(1-\bar w/d)=a^2x(1-by)\cr
\bar y(1-\bar x/a)=b^2y(1-cz)\cr
\bar z(1-\bar y/b)=c^2z(1-dw)\cr
\bar w(1-\bar z/c)=d^2w(1-ax)\cr},
\end{equation}
and is solved by
\begin{equation}
\left(\matrix{\bar x\cr\bar y\cr\bar z\cr\bar w\cr}\right)
=\cases{
\left(\matrix{a^2x\ds{{1-by-cz+bcyz+cdzw\over 1-cz-dw+adxw+cdzw}}\cr
              b^2y\ds{{1-cz-dw+adxw+cdzw\over 1-ax-dw+abxy+adxw}}\cr
              c^2z\ds{{1-ax-dw+abxy+adxw\over 1-ax-by+abxy+bcyz}}\cr
              d^2w\ds{{1-ax-by+abxy+bcyz\over 1-by-cz+bcyz+cdzw}}\cr}
              \right)\qquad :{\rm A-type}\cr
\ \ \cr
\ \ \cr
\left(\matrix{a(1-by)\cr b(1-cz)\cr c(1-dw)\cr d(1-ax)\cr}\right)\qquad :{\rm B-type}.\cr}
\end{equation}

\section{Further generalization}

Let us further generalize (\ref{3dGLV}) to
\begin{equation}
\cases{
\bar x(W-\bar z/c)=aX,\cr
\bar y(U-\bar x/a)=bY,\cr
\bar z(V-\bar y/b)=cZ,\cr}
\label{3dGGLV}
\end{equation}
where $U,V,W$ are some functions of $x,y,z$. (\ref{square root}) is then changed to
\begin{equation}
\bar x=a{UVW-WY-UZ+VX\pm\sqrt{(UVW-WY-UZ-VX)^2-4XYZ}\over 2(VW-Z)}.
\label{square root 2}
\end{equation}
We want to know $X,Y,Z$ such that the expression under the square root in (\ref{square root 2}) becomes a perfect square. We find that
\begin{equation}
\cases{
\bar x(W-\bar z/c)=a^2x(U-by),\cr
\bar y(U-\bar x/a)=b^2y(V-cz),\cr
\bar z(V-\bar y/b)=c^2z(W-ax),\cr}
\label{X=x(U-y)}
\end{equation}
and two other equations, obtained from (\ref{X=x(U-y)}) by cyclic permutations (\ref{cyclic permutation}), do satisfy the requirement. Solving (\ref{X=x(U-y)}) for $(\bar x,\bar y,\bar z)$ we obtain two maps
\begin{equation}
\left(\matrix{\bar x\cr \bar y\cr \bar z\cr}\right)
=
\cases{
\left(\matrix{a^2x\ds{{UV-bVy+bcyz\over VW-cWz+acxz}}\cr 
b^2y\ds{{VW-cWz+acxz\over WU-aUx+abxy}}\cr c^2z\ds{{WU-aUx+abxy\over UV-bVy+bcyz}}\cr}\right)
\qquad :{\rm A-type}
\cr
\ \ \ \ 
\cr
\ \ \ \ 
\cr
\left(\matrix{a(U-by)\cr b(V-cz)\cr c(W-ax)\cr}\right)\qquad :{\rm B-type}.
}
\label{solutions to X=x(U-y)}
\end{equation}

The maps of (\ref{solutions to X=x(U-y)}) have constants
\begin{equation}
R:={\bar x\bar y\bar z\over abcxyz},\qquad 
S:={(U-by)(V-cz)(W-ax)\over (aU-\bar x)(bV-\bar y)(cW-\bar z)}
\label{RS}
\end{equation}
under the A-type map and
$$
{(aU-\bar x)(bV-\bar y)(cW-\bar z)\over abcxyz},\qquad
{\bar x\bar y\bar z\over (U-by)(V-cz)(W-ax)}
$$
under the B-type map. Note that the B-type map exchanges $R\leftrightarrow S$.

When $(U,V,W)$ are functions of $(x,y,z)$, the time reversal symmetry is lost, hence $(x,y,z)$ obtained by solving (\ref{solutions to X=x(U-y)}) will not be rational polynomials of $(\bar x,\bar y,\bar z)$ in general. The symmetry is recovered if $U,V,W$ have some symmetry under the exchange
$(\bar x,\bar y,\bar z)\leftrightarrow (x,y,z)$. Apparently the $q$-difference Lotka-Volterra system satisfies this condition. We'll present other examples in what follows.

\subsection{$q$-KdV maps}

We have noted that the A-type map of (\ref{solutions to X=x(U-y)}) preserves $R$, irrespective  of $U,V,W$. Thus if $U,V,W$ are given by 
\begin{equation}
U=V=W=abcxyz
\label{U=V=W=abcxyz}
\end{equation}
the requirement is fulfilled. Namely we can substitute another expression
\begin{equation}
U=V=W=\bar x\bar y\bar z/abc
\label{U=V=W=barabcxyz}
\end{equation}
of $U,V,W$ to the left hand side of (\ref{X=x(U-y)}) to solve the inverse map.

In fact a substitution of (\ref{U=V=W=abcxyz}) into the A-type map of (\ref{solutions to X=x(U-y)}) yields
\begin{equation}
\left(\matrix{\bar x\cr \bar y\cr \bar z\cr}\right)
=
\left(\matrix{aby\ds{{1-abxy+a^2bcx^2yz\over 1-bcyz+ab^2cxy^2z}}\cr 
bcz\ds{{1-bcyz+ab^2cxy^2z\over 1-cazx+abc^2xyz^2}}\cr 
cax\ds{{1-cazx+abc^2xyz^2\over 1-abxy+a^2bcx^2yz}}\cr}\right).
\label{A-type of 27}
\end{equation}
We can solve this map inversely for $(x,y,z)$ and obtain
$$
\left(\matrix{ x\cr y\cr z\cr}\right)
=
\left(\matrix{\ds{{\bar z\over a^2}}\ds{{a^2bc-ab\bar x\bar z+\bar x^2\bar y\bar z\over abc^2-ac\bar y\bar z+\bar x\bar y\bar z^2}}\cr 
\ds{{\bar x\over b^2}}\ds{{ab^2c-bc\bar x\bar y+\bar x\bar y^2\bar z\over a^2bc-ab\bar x\bar z+\bar x^2\bar y\bar z}}\cr \ds{{\bar y\over c^2}}\ds{{abc^2-ac\bar y\bar z+\bar x\bar y\bar z^2\over ab^2c-bc\bar x\bar y+\bar x\bar y^2\bar z}}\cr}\right).
$$
It is not difficult to convince ourselves that the same map is obtained by solving (\ref{X=x(U-y)}) for $(x,y,z)$ after the substitution of (\ref{U=V=W=barabcxyz}).

Now we like to show a relation of (\ref{A-type of 27}) with the discrete KdV map. The latter has been proposed in the form\cite{Hirota, HT}
\begin{equation}
\cases{
\ds{{\bar x-{1\over\bar y}=x-{1\over z}}},\cr
\ \ \cr
\ds{{\bar y-{1\over\bar z}=y-{1\over x}}},\cr
\ \ \cr
\ds{{\bar z-{1\over\bar x}=z-{1\over y}}},\cr}
\label{d-KdV}
\end{equation}
A direct substitution of
\begin{equation}
\left(\matrix{\bar x\cr \bar y\cr \bar z\cr}\right)
=
\left(\matrix{y\ds{{1-xy+x^2yz\over 1-yz+xy^2z}}\cr 
z\ds{{1-yz+xy^2z\over 1-zx+xyz^2}}\cr 
x\ds{{1-zx+xyz^2\over 1-xy+x^2yz}}\cr}\right)
\label{d-KdV solution}
\end{equation}
will show that this solves (\ref{d-KdV}). But (\ref{d-KdV solution}) is simply a special case of (\ref{A-type of 27}) with $a=b=c=1$. From this reason we call 
\begin{equation}
\cases{
\ds{{{\bar x\over a}-{b\over\bar y}=ax-{1\over cz}}},\cr
\ \ \cr
\ds{{{\bar y\over b}-{c\over\bar z}=by-{1\over ax}}},\cr
\ \ \cr
\ds{{{\bar z\over c}-{a\over\bar x}=cz-{1\over by}}},\cr}
\label{qX=x(U-y)}
\end{equation}
the $q$-KdV map, which is obtained from (\ref{X=x(U-y)}) after substitutions of (\ref{U=V=W=abcxyz}) and (\ref{U=V=W=barabcxyz}).

\subsection{$q$-Toda maps}

In addition to $x,y,z$ we introduce variables $u,v,w$ and assume for $U,V,W$ in (\ref{X=x(U-y)}) the following relations:
\begin{eqnarray}
U&=&by+a'u={\bar x\over a}+{\bar v\over b'},\nonumber\\
V&=&cz+b'v={\bar y\over b}+{\bar w\over c'},
\label{equation 1 for u,v,w}\\
W&=&ax+c'w={\bar z\over c}+{\bar u\over a'}.\nonumber
\end{eqnarray}
Then we find solutions to (\ref{X=x(U-y)})
\begin{equation}
\left(\matrix{\bar x\cr \bar y\cr\bar z\cr\bar u\cr\bar v\cr\bar w\cr}\right)=
\cases{
\left(\matrix{a^2x\ds{{a'b'uv+ca'uz+bcyz\over b'c'vw+ab'vx+acxz}}\cr b^2y\ds{{b'c'vw+ab'vx+acxz\over a'c'uw+bc'wy+abxy}}\cr c^2z\ds{{a'c'uw+bc'wy+abxy\over a'b'uv+a'cuz+bcyz}}\cr
a'^2u\ds{{b'c'vw+b'avx+acxz\over a'b'uv+a'cuz+bcyz}}\cr
b'^2v\ds{{a'c'uw+bc'wy+abxy\over b'c'vw+b'avx+acxz}}\cr
c'^2w\ds{{a'b'uv+a'cuz+bcyz\over a'c'uw+bc'wy+abxy}}\cr}\right)\qquad :{\rm A-type}
\cr
\ \ \cr
\ \ \cr
\left(\matrix{aa'u\cr bb'v\cr cc'w\cr aa'x\cr bb'y\cr cc'z\cr}\right)\qquad :{\rm B-type}.
}
\end{equation}

Note that this system is equivalent to the equations
$$
by+a'u={\bar x\over a}+{\bar v\over b'},\quad
cz+b'v={\bar y\over b}+{\bar w\over c'},\quad
ax+c'w={\bar z\over c}+{\bar u\over a'}
$$
\begin{equation}
{\bar x\bar u\over aa'}=aa'xu,\qquad
{\bar y\bar v\over bb'}=bb'yv,\qquad
{\bar z\bar w\over cc'}=cc'zw,
\label{simple Toda}
\end{equation}
which is nothing but the discrete time Toda equations\cite{HT} when $a=b=c=a'=b'=c'=1$. Hence we obtained a $q$ difference discrete time Toda system.

\subsection{$q$-KP maps}

We now consider three sets of equations of the type (\ref{X=x(U-y)}):
\begin{equation}
\cases{
\bar x_m(W_m-\bar z_m/c_m)=a_m^2x_m(U_m-b_my_m),\cr
\bar y_m(U_m-\bar x_m/a_m)=b_m^2y_m(V_m-c_mz_m),\qquad m=1,2,3\cr
\bar z_m(V_m-\bar y_m/b_m)=c_m^2z_m(W_m-a_mx_m)\cr}.
\label{3KP}
\end{equation}

By introducing new variables $(u_m,v_m,w_m)$ and $(\bar u_m,\bar v_m,\bar w_m)$ we assume for $(U,V,W)$ the following form
\begin{equation}
\cases{
\ds{U_m={\bar x_m\over a_m}+{\bar v_m\over b'_m}=b_my_m+a'_{m+1}u_{m+1},}\cr
\ds{V_m={\bar y_m\over b_m}+{\bar w_m\over c'_m}=c_mz_m+b'_{m+1}v_{m+1}},\qquad m=1,2,3
\cr
\ds{W_m={\bar z_m\over c_m}+{\bar u_m\over a'_m}=a_mx_m+c'_{m+1}w_{m+1}}.\cr}
\label{U_m=}
\end{equation}
Here the periodic boundary conditions $x_{m+3}=x_m, u_{m+3}=u_m, a_{m+3}=a_m$ etc. are used. Solving (\ref{3KP}) for $(\bar x_m,\bar y_m,\bar z_m,\bar u_m,\bar v_m,\bar w_m)$ we find
\begin{equation}
\left(\matrix{\bar x_m\cr\bar y_m\cr\bar z_m\cr \bar u_m\cr\bar v_m\cr\bar w_m\cr}\right)
=\cases{
\left(\matrix{a_m^2x_m\ds{{a'_{m+1}b'_{m+1}u_{m+1}v_{m+1}+a'_{m+1}c_mu_{m+1}z_m+b_mc_my_mz_m\over b'_{m+1}c'_{m+1}v_{m+1}w_{m+1}+b'_{m+1}a_mv_{m+1}x_m+a_mc_mx_mz_m}}\cr
b_m^2y_m\ds{{b'_{m+1}c'_{m+1}v_{m+1}w_{m+1}+b'_{m+1}a_mv_{m+1}x_m+c_ma_mz_mx_m\over c'_{m+1}a'_{m+1}w_{m+1}u_{m+1}+c'_{m+1}b_mw_{m+1}y_m+b_ma_mx_my_m}}\cr
c_m^2z_m\ds{{c'_{m+1}a'_{m+1}w_{m+1}u_{m+1}+c'_{m+1}b_mw_{m+1}y_m+a_mb_mx_my_m\over a'_{m+1}b'_{m+1}u_{m+1}v_{m+1}+a'_{m+1}c_mu_{m+1}z_m+b_mc_my_mz_m}}\cr
a'_{m+1}a'_mu_{m+1}\ds{{b'_{m+1}c'_{m+1}v_{m+1}w_{m+1}+b'_{m+1}a_mv_{m+1}x_m+a_mc_mx_mz_m\over a'_{m+1}b'_{m+1}u_{m+1}v_{m+1}+a'_{m+1}c_mu_{m+1}z_m+b_mc_my_mz_m}}\cr
b'_{m+1}b'_mv_{m+1}\ds{{c'_{m+1}a'_{m+1}w_{m+1}u_{m+1}+c'_{m+1}b_mw_{m+1}y_m+a_mb_mx_my_m\over b'_{m+1}c'_{m+1}v_{m+1}w_{m+1}+b'_{m+1}a_mv_{m+1}x_m+c_ma_mz_mx_m}}\cr
c'_{m+1}c'_mw_{m+1}\ds{{a'_{m+1}b'_{m+1}u_{m+1}v_{m+1}+a'_{m+1}c_mu_{m+1}z_m+b_mc_my_mz_m\over c'_{m+1}a'_{m+1}w_{m+1}u_{m+1}+c'_{m+1}b_mw_{m+1}y_m+a_mb_mx_my_m}}\cr
}\right)
\cr
\ \ \ 
\cr
\qquad \qquad\qquad\qquad\qquad\qquad\qquad:{\rm A}_m{\rm -type}
\cr
\ \ \ 
\cr
\left(\matrix{a_ma'_{m+1}u_{m+1}\cr b_mb'_{m+1}v_{m+1}\cr c_mc'_{m+1}w_{m+1}\cr a_ma'_mx_m\cr b_mb'_my_m\cr c_mc'_mz_m\cr}\right)\qquad :{\rm B}_m{\rm -type}
}
\end{equation}
with $m=1,2,3$. All eight combinations of the A$_m$-type and B$_m$-type maps
$$
(A_1,A_2,A_3),\  (A_1,A_2,B_3),\cdots, (B_1,B_2,B_3)
$$
are possible map.

We would like to note here that the set of equations (\ref{3KP}) together with (\ref{U_m=}) are $q$-difference discrete time KP-equation or the discrete 2D Toda equation. In fact the substitution of (\ref{U_m=}) into (\ref{3KP}) shows that the set of equations (\ref{3KP}) and (\ref{U_m=}) are equivalent to
\begin{equation}
{\bar x_{mn}\bar u_{mn}\over a_{mn}b_{mn}}=a_{m,n}b_{m+1,n}x_{m,n}u_{m+1,n},
\label{q-d-KP 2}
\end{equation}
\begin{equation}
{\bar x_{mn}\over a_{mn}}+{\bar u_{m,n+1}\over b_{m,n+1}}=a_{m,n+1}x_{m,n+1}+b_{m+1,n}u_{m+1,n},
\label{q-d-KP 1}
\end{equation}
respectively. Here we use the notations
$$
(x_{m,1}, x_{m,2}, x_{m,3})=(x_m,y_m,z_m),\quad
(u_{m,1}, u_{m,2}, u_{m,3})=(u_m,v_m,w_m),
$$
$$
(a_{m,1}, a_{m,2}, a_{m,3})=(a_m,b_m,c_m),\quad
(b_{m,1}, b_{m,2}, b_{m,3})=(a'_m,b'_m,c'_m).
$$

The discrete KP (or discrete 2D Toda) equation has been given by\cite{HTI}
$$
V_\nu (\xi +1,\eta )I_\nu (\xi ,\eta +1)=V_\nu (\xi ,\eta )I_{\nu +1}(\xi ,\eta ),
$$
$$
V_{\nu -1}(\xi +1,\eta )+I_\nu (\xi ,\eta +1)=V_\nu (\xi ,\eta )+I_{\nu }(\xi ,\eta ).
$$
The correspondence will be established if we put
$$
x^j_{mn}=V_\nu (\xi ,\eta ),\qquad u^{j+1}_{mn}=I_\nu (\xi ,\eta +1)
$$
in (\ref{q-d-KP 2}) and (\ref{q-d-KP 1}) and identify the variables according to 
$$
\nu =m+n,\qquad \xi =j+m,\qquad \eta =-m.
$$
Therefore (\ref{3KP}) together with (\ref{U_m=}) is a $q$-differnce discrete KP equation.

\section{Discussions}

We have found a large class of maps which share the same properties with those of discrete Lotka-Vorterra map. All of them have two types of map, which we called A-type and B-type. The B-type map does not create a new orbit, but exchanges orbits generated by the A-type map. 

In the rest of this paper we'll discuss about conservation laws in our system. We already pointed out in the case of dLV that $C$ and $D$ of (\ref{C}) and (\ref{D}) are conserved along the sequence of the map. If we restrict a map to the A-type there are additional constants $r$ and $s$ given by (\ref{r,s}). 

This is true also in other cases if $a=b=c=1$ and $U=V=W$. In fact from (\ref{X=x(U-y)}) we see that
$$
C_U=(x+y+z)U-xy-yz-zx,\qquad D_U=xyz(U-x)(U-y)(U-z)
$$
are constants. The dLV and dKdV belong to this case. If we consider only A-type map and assume $a=b=c=1$, (\ref{solutions to X=x(U-y)}) shows that $r=xyz$ is a constant of the map irrespective to $U,V,W$.

The discrete time Toda lattice contains more variables. We can read off conserved quantities from (\ref{simple Toda}) as
$$
u+v+w+x+y+z,\qquad ux,\qquad vy,\qquad wz
$$
if $a=b=c=a'=b'=c'=1$. In the case of A-type of map 
$$
xyz,\qquad uvw
$$
are also conserved as long as $a=b=c=a'=b'=c'=1$.

When the maps are generalized by introducing the parameters $a,b,c,\cdots$, these quantities are not conserved along the map anymore. For instance we find, in the case of A-type map of (\ref{solutions to X=x(U-y)}), $r=xyz$ changes according to
\begin{equation}
\bar x\bar y\bar z=q^2xyz,\qquad q=abc.
\end{equation}
This means that the scale is changed by an equal rate $q^2$ at every step of the map. Hence the quantity $xyz$ is not constant.

The scale change of this form was identified with a time evolution of the system in ref\cite{KNY}. From this point of view we can interprete $2\ln q$ as a velocity of the variable $\ln (xyz)$. Namely, writing
$$
r(t)=xyz=q^{2t}r(0),
$$
we have
$$
{d\ln r(t)\over dt}=2\ln q.
$$
As we already know this is a constant of the map of A-type. In the special case of $q=1$, the velocity is zero, hence $r(t)$ itself turns out to be a constant. Therefore it is natural to understand that the quantity $R$ in (\ref{RS}) is a conserved quantity of the A-type equation of motion of (\ref{solutions to X=x(U-y)}). $\ln (qR)$ is a difference of variables at two subsequent steps of the map. One may interprete it as a center of mass velocity of the coordinates $(\ln x, \ln y, \ln z)$:
$$
\ln (qR)=(\ln \bar x+\ln \bar y+\ln \bar z)-(\ln x+\ln y+\ln z)={d\over dt}(\ln x+\ln y+\ln z)
$$

In addition to $R$ we have another quantity $S$ which remains constant under the map. $S$, which is given by (\ref{RS}), also depends on variables of two steps of the map. It has no direct interpretation like $R$, but becomes $R$ by the B-type map.



\begin{thebibliography}{99}
\bibitem{KNY} K.Kajiwara, M.Noumi and Y.Yamada, `` A Study on the fourth $q$-Painlev\'e Equation", nlin. SI/0012063,

T.Masuda, `` On the Rational Solutions of $q$-Painlev\'e V Equation", nlin.SI/0107050.
\bibitem{Hirota} 
R.Hirota, J.Phys. Soc. Jpn. {\bf 43} (1977) 1424.
\bibitem{Hirota2}
R.Hirota, J. Phys. Soc. Jpn. {\bf 50} (1981) 3785.
\bibitem{JM}
M.Jimbo and T.Miwa, J. Phys. Soc. Jpn. {\bf 51} (1982) 4116, 4125. 
\bibitem{FN}
N.C.Freeman and J.J.C.Nimmo, Phys. Lett. {\bf 95A} (1983) 1.
\bibitem{QRT}
G.Quispel, J.Roberts and C.Thompson, Phys. Lett. {\bf A126} (1988) 419.
\bibitem{GRP}
B.Grammaticos, A.Ramani and J.Hietarinta, Phys. Rev. Lett. {\bf 67} (1991) 1829.
\bibitem{NP}
F.Nijhoff and V.G.Papageorgiou, Phys. Lett. {\bf 153A} (1991) 337.
\bibitem{JBH}
N.Joshi, D.Burtonclay and R.G.Halburd, Lett. Math. Phys. {\bf 26} (1992) 123
\bibitem{HTI}
R.Hirota, S.Tsujimoto and T.Imai, ``{\it Difference Scheme of Soliton Equations}", in {\it Future Directions of Nonlinear Dynamics in Physical and Biological Systems}, ed. by P.L.Christiansen at al., (Plenum Press, New York, 1993) p.7. 
\bibitem{OHT}
Y.Ohta, R.Hirota and S.Tsujimoto, J.Phys. Soc. Jpn. {\bf 62} (1993) 1872.
\bibitem{HT} 
R.Hirota, and S.Tsujimoto, J.Phys.Soc.Jpn. {\bf 64} (1995) 3125-3127
\bibitem{Kor}
I.G.Korepanov, ``Algebraic integrable dynamical systems, 2+1-dimensional models in wholly discrete space-time, and
     inhomogeneous models in 2-dimensional statistical physics", solv-int/9506003,

R.M.Kashaev, ``On Discrete 3-Dimensional Equations Associated with the Local Yang-Baxter Relation" solv-int/9512005.

\bibitem{CM}
R.Conte and M.Musette, in ``{\it Theory of nonlinear special functions : the Painlev\'e transcendents}", eds. L. Vinet and P. Winternitz (Springer, Berlin, 1998). Proceedings of Montreal, 13--17 May 1996

\bibitem{Kup}
B.Kupershmidt, {\it KP or mKP} Mathematical Surveys and Monographs Vol {\bf 78} (American Mathematical Society, 2000) and references therein.
\bibitem{Su} 
Y.Suris, see, for instance, arXiv:solv-int/9902003.
\bibitem{HV}
J.Hietarinta and C.Viallet, Phys. Rev. Lett. {\bf 81} (1998) 325, 

S.Lafortune, A.Ramani, B.Grammaticos, Y.Ohta and K.M.Tamizhmani, arXiv:nlin.SI/0104020

\bibitem{NSSY} 
Y.Narita, S.Saito, N.Saitoh and K.Yoshida, J. Phys. Soc. Jpn. {\bf 70} (2001) 1246.

\end{thebibliography}
\end{document}